\documentclass[pra,aps,nopacs,twocolumn,twoside,superscriptaddress]{revtex4}

\usepackage{amsmath,amsfonts,amssymb,caption,color,epsfig,graphics,graphicx,hyperref,latexsym,mathrsfs,revsymb,theorem,url,verbatim,epstopdf,mathtools,enumerate}
\usepackage{subfigure}\usepackage{makecell}\usepackage{multirow}\usepackage{diagbox}\usepackage{xcolor}
\usepackage[most]{tcolorbox}
\hypersetup{colorlinks,linkcolor={blue},citecolor={blue},urlcolor={blue}}

\usepackage[qm]{qcircuit}

\newtheorem{definition}{Definition}
\newtheorem{proposition}[definition]{Proposition}
\newtheorem{lemma}[definition]{Lemma}

\newtheorem{theorem}[definition]{Theorem}
\newtheorem{corollary}[definition]{Corollary}
\newtheorem{conjecture}[definition]{Conjecture}

\newtheorem{remark}[definition]{Remark}
\newtheorem{example}[definition]{Example}
\newtheorem{question}[definition]{Question}
\newtheorem{memo}[definition]{Memo}

\def\squareforqed{\hbox{\rlap{$\sqcap$}$\sqcup$}}
\def\qed{\ifmmode\squareforqed\else{\unskip\nobreak\hfil
\penalty50\hskip1em\null\nobreak\hfil\squareforqed
\parfillskip=0pt\finalhyphendemerits=0\endgraf}\fi}
\def\endenv{\ifmmode\;\else{\unskip\nobreak\hfil
\penalty50\hskip1em\null\nobreak\hfil\;
\parfillskip=0pt\finalhyphendemerits=0\endgraf}\fi}
\newenvironment{proof}{\noindent \textbf{{Proof.~} }}{\qed}
\def\Dbar{\leavevmode\lower.6ex\hbox to 0pt
{\hskip-.23ex\accent"16\hss}D}
\makeatletter
\def\url@leostyle{%
  \@ifundefined{selectfont}{\def\UrlFont{\sf}}{\def\UrlFont{\small\ttfamily}}}
\makeatother
\def\bcj{\begin{conjecture}}
\def\ecj{\end{conjecture}}
\def\bcr{\begin{corollary}}
\def\ecr{\end{corollary}}
\def\bd{\begin{definition}}
\def\ed{\end{definition}}
\def\bea{\begin{eqnarray}}
\def\eea{\end{eqnarray}}
\def\beq{\begin{equation}}
\def\eeq{\end{equation}}
\def\bal{\begin{aligned}}
\def\eal{\end{aligned}}
\def\bem{\begin{enumerate}}
\def\eem{\end{enumerate}}
\def\bex{\begin{example}}
\def\eex{\end{example}}
\def\bim{\begin{itemize}}
\def\eim{\end{itemize}}
\def\bl{\begin{lemma}}
\def\el{\end{lemma}}
\def\bma{\begin{bmatrix}}
\def\ema{\end{bmatrix}}
\def\bpf{\begin{proof}}
\def\epf{\end{proof}}
\def\bpp{\begin{proposition}}
\def\epp{\end{proposition}}
\def\bqu{\begin{question}}
\def\equ{\end{question}}
\def\br{\begin{remark}}
\def\er{\end{remark}}
\def\bt{\begin{theorem}}
\def\et{\end{theorem}}
\def\bmm{\begin{memo}}
\def\emm{\end{memo}}

\def\btb{\begin{tabular}}
\def\etb{\end{tabular}}

\newcommand{\nc}{\newcommand}

\def\a{\alpha}
\def\b{\beta}
\def\g{\gamma}

\def\t{\theta}

\def\r{\rho}
\def\s{\sigma}

\def\ph{\varphi}

\def\ps{\psi}
\def\og{\omega}

\def\L{\Lambda}

\nc{\bbA}{\mathbb{A}} \nc{\bbB}{\mathbb{B}} \nc{\bbC}{\mathbb{C}}
 \nc{\bbD}{\mathbb{D}} \nc{\bbE}{\mathbb{E}} \nc{\bbF}{\mathbb{F}}
 \nc{\bbG}{\mathbb{G}} \nc{\bbH}{\mathbb{H}} \nc{\bbI}{\mathbb{I}}
 \nc{\bbJ}{\mathbb{J}} \nc{\bbK}{\mathbb{K}} \nc{\bbL}{\mathbb{L}}
 \nc{\bbM}{\mathbb{M}} \nc{\bbN}{\mathbb{N}} \nc{\bbO}{\mathbb{O}}
 \nc{\bbP}{\mathbb{P}} \nc{\bbQ}{\mathbb{Q}} \nc{\bbR}{\mathbb{R}}
 \nc{\bbS}{\mathbb{S}} \nc{\bbT}{\mathbb{T}} \nc{\bbU}{\mathbb{U}}
 \nc{\bbV}{\mathbb{V}} \nc{\bbW}{\mathbb{W}} \nc{\bbX}{\mathbb{X}}
 \nc{\bbZ}{\mathbb{Z}}

 \nc{\bA}{{\bf A}} \nc{\bB}{{\bf B}} \nc{\bC}{{\bf C}}
 \nc{\bD}{{\bf D}} \nc{\bE}{{\bf E}} \nc{\bF}{{\bf F}}
 \nc{\bG}{{\bf G}} \nc{\bH}{{\bf H}} \nc{\bI}{{\bf I}}
 \nc{\bJ}{{\bf J}} \nc{\bK}{{\bf K}} \nc{\bL}{{\bf L}}
 \nc{\bM}{{\bf M}} \nc{\bN}{{\bf N}} \nc{\bO}{{\bf O}}
 \nc{\bP}{{\bf P}} \nc{\bQ}{{\bf Q}} \nc{\bR}{{\bf R}}
 \nc{\bS}{{\bf S}} \nc{\bT}{{\bf T}} \nc{\bU}{{\bf U}}
 \nc{\bV}{{\bf V}} \nc{\bW}{{\bf W}} \nc{\bX}{{\bf X}}
 \nc{\bZ}{{\bf Z}}

\nc{\cA}{{\cal A}} \nc{\cB}{{\cal B}} \nc{\cC}{{\cal C}}
\nc{\cD}{{\cal D}} \nc{\cE}{{\cal E}} \nc{\cF}{{\cal F}}
\nc{\cG}{{\cal G}} \nc{\cH}{{\cal H}} \nc{\cI}{{\cal I}}
\nc{\cJ}{{\cal J}} \nc{\cK}{{\cal K}} \nc{\cL}{{\cal L}}
\nc{\cM}{{\cal M}} \nc{\cN}{{\cal N}} \nc{\cO}{{\cal O}}
\nc{\cP}{{\cal P}} \nc{\cQ}{{\cal Q}} \nc{\cR}{{\cal R}}
\nc{\cS}{{\cal S}} \nc{\cT}{{\cal T}} \nc{\cU}{{\cal U}}
\nc{\cV}{{\cal V}} \nc{\cW}{{\cal W}} \nc{\cX}{{\cal X}}
\nc{\cZ}{{\cal Z}}

\nc{\hA}{{\hat{A}}} \nc{\hB}{{\hat{B}}} \nc{\hC}{{\hat{C}}}
\nc{\hD}{{\hat{D}}} \nc{\hE}{{\hat{E}}} \nc{\hF}{{\hat{F}}}
\nc{\hG}{{\hat{G}}} \nc{\hH}{{\hat{H}}} \nc{\hI}{{\hat{I}}}
\nc{\hJ}{{\hat{J}}} \nc{\hK}{{\hat{K}}} \nc{\hL}{{\hat{L}}}
\nc{\hM}{{\hat{M}}} \nc{\hN}{{\hat{N}}} \nc{\hO}{{\hat{O}}}
\nc{\hP}{{\hat{P}}} \nc{\hR}{{\hat{R}}} \nc{\hS}{{\hat{S}}}
\nc{\hT}{{\hat{T}}} \nc{\hU}{{\hat{U}}} \nc{\hV}{{\hat{V}}}
\nc{\hW}{{\hat{W}}} \nc{\hX}{{\hat{X}}} \nc{\hZ}{{\hat{Z}}}

\nc{\hn}{{\hat{n}}}

\def\max{\mathop{\rm max}}

\def\tr{\mathop{\rm Tr}}

\def\dg{\dagger}

\newcommand{\ket}[1]{|#1\rangle}
\newcommand{\proj}[1]{| #1\rangle\!\langle #1 |}

\newcommand{\braket}[2]{\langle#1|#2\rangle}

\newcommand{\norm}[1]{\lVert#1\rVert}
\newcommand{\abs}[1]{|#1|}

\def\Dbar{\leavevmode\lower.6ex\hbox to 0pt
{\hskip-.23ex\accent"16\hss}D}

\begin{document}

\normalsize

\title{Device-independent quantum state discrimination}

\date{\today}
\author{Xinyu Qiu}\email[]{xinyuqiu@buaa.edu.cn}
\affiliation{LMIB(Beihang University), Ministry of education, and School of Mathematical Sciences, Beihang University, Beijing 100191, China}
\author{Lin Chen}\email[]{linchen@buaa.edu.cn (corresponding author)}
\affiliation{LMIB(Beihang University), Ministry of education, and School of Mathematical Sciences, Beihang University, Beijing 100191, China}
\affiliation{International Research Institute for Multidisciplinary Science, Beihang University, Beijing 100191, China}

\begin{abstract}
Quantum state discrimination depicts the general progress of extracting classical information from quantum systems. We show that quantum state discrimination can be realized in a device-independent scenario using tools of self-testing results. That is, the states can be discriminated credibly with the untrusted experiment devices by the correspondence between quantum correlations and states. In detail, we show that two arbitrary states can be discriminated in a device-independent manner when they are not conjugate with each other, while other states can be discriminated measurement-device-independently. To fulfill the device-independent requirement, the measurements are restricted on Pauli observables. The influence of this restriction is acceptable based on the guessing probability analysis for minimum error discrimination.
\end{abstract}

\maketitle

\section{Introduction}
Quantum state discrimination is a fundamental topic in quantum information theory. Starting with the attempts to formalize information
processing of physical systems, it depicts a characteristic feature of quantum mechanics,  that is, the impossibility to distinguish nonorthogonal states perfectly \cite{stephen2009quantum}. This feature has enabled quantum advantages in various applications such as quantum cryptography \cite{Bennett1992quantum, van2002Unambiguous, gisin2002quantum}, communication \cite{bennett1992communication, bennett1993teleport} and others \cite{brunner2013dimension, konig2009operational}.
Quantum state discrimination is conventionally  scenarized as a game involving two parties, Alice and Bob, who agree on a set of states with prior probabilities known publicly. Alice encodes the information on the states and transfers the state to Bob who realizes discrimination by trusted measurements. Obviously, the conventional state discrimination relies on the precise knowledge of the measurements, and thus it is described as device-dependent.  If the physical system used for discrimination is poorly characterized or highly complex, the state discrimination may be false or unaccessible.

One possible solution for this problem is derived from the device-independent (DI) information processing. Since it is reasonable to trust a source device than a detection device, the measurement device independence (MDI) allows for the dependence on the function of measuring devices \cite{Graffitti2020measurement, xu2014implement}. As a stronger form, device independence allows one to simply treat all devices as black boxes, and certify the physical properties of quantum systems without requiring precise knowledge of the underlying physics \cite{n2016device,lee2018towards}. Self-testing pushes the DI certification to its strongest form \cite{yao2004self, self2020ivan}.  It allows to infer the underlying physics of a quantum experiment only from the observed statistics. The maximum violation of some Bell inequalities implies that the  states and measurements can be uniquely identified up to local isometry.
Recently, it has been shown that all entangled states can be self tested \cite{ivan2023quantum}. The precise form of the quantum measurements can also be certified, such as Pauli observables \cite{bowles2018pra} and Bell measurements \cite{Renou2018self}.  

Self-testing offers prominent progress for the quantum system certification experiments \cite{Shalm2021device}, as well as various DI protocols, such as DI quantum cryptography \cite{mayers1998quantum,jedrzej2016di}, random generation \cite{acin2016certified}, and entanglement certification \cite{bowles2018prl}.    Quantum state discrimination essentially describes the general process of extracting classical information from quantum systems and underlies various applications in  information processing tasks.  Realizing it in a DI scenario will naturally show the following practical advantages.  First, one can realize the state discrimination without requiring precise knowledge of the underlying physics.  Second, the trust of devices can be removed and thus the security is improved.  Third, the experimental errors can be treated at the level of observed statistics.

In a DI scenario, the quantum state discrimination should be described as follows. 
 Alice and Bob do not know the quantum mechanism and experimental setup. Bob still wishes to obtain which of the state is transmitted from the observed statistics only.  All they know about the system is that the  state is  from the ensemble $\{q_k,\ket{\psi_k}\}$. This can hardly be accomplished by self-testing  states, as it relies on Bell nonlocality that forbids 
to certify local properties of states. Besides, in the self-testing all parties know their target state and just want to certify whether it is as their prediction. Here in the quantum state discrimination, we have no idea which state is and aim to figure it out. 

In this paper, we introduce the device-independent quantum state discrimination protocol. We aim to turn a conventional quantum state discrimination, where measurements are characterized, into a DI state discrimination protocol by removing the assumption that the parties know the precise form of the measurements. In a quantum network in FIG. \ref{fig:network1}, all devices are modeled as black boxes that take an input and returns an output. Using the tools of self-testing, the measurements and auxiliary states can be certified. On the basis of that, 
 the  states are discriminated credibly with the untrusted experiment devices by the correspondence between quantum correlations and states.  Two arbitrary states can be discriminated in a DI manner when they are not conjugate with each other. Since the DI assumptions only allow to characterize the measurements up to configuration, the states conjugate with each other can only be discriminated in a MDI manner. In order to fulfill the DI requirement,  measurements in this protocol are restricted as Pauli observables. The guessing probability for minimum error discrimination is analyzed. Taking all real states as an example, the guessing probability between Pauli measurement strategy and the optimal one is compared. It is shown that the bias of guessing probability is less than 15\% on maximum and 0.33\% on average over all possible states. So the influence of restriction is acceptable.

The rest of this paper is organized as follows. In Sec. \ref{sec:pre,ST}, we show some existing results for self-testing, which will be used in our protocol. In Sec. \ref{sec:discrimination}, we show the  DI discrimination of states in a quantum network. In Sec. \ref{sec: mainguessing probability}, we present the guessing probability analysis for this protocol in minimum-error discrimination. We conclude in Sec. \ref{sec:conclusion}.

\section{Preliminaries: self testing}
\label{sec:pre,ST}
The scenario of self-testing is depicted as follows. The action of the parties, Alice and Bob, is described by measuring the local binary observable $M_j=M_{+1|j}-M_{-1|j}$, where  $M_{k|j}$ is the measurement projectors associated with outcome (output) $k$ for the chosen measurement operator (input) $j$. The measurement operators for Alice and Bob is labeled by $x,y$ with the outcomes $a,b$. By collecting the statistics for inputs and outputs for several rounds, the quantum correlations are derived as
\begin{eqnarray}
	\label{def:p(abxy)}
p(a,b|x,y)=\tr[M_{a|x}\otimes M_{b|y}\proj{\ps}^{AB}].
\end{eqnarray}    

A key ingredient of self-testing is  the Bell inequalities. Specifically, we consider the CHSH inequalities, which are examples of a larger set known as Bell inequalities. They are defined by the expression \cite{clauser1969proposed}
\begin{eqnarray}
	\label{eq:cI}
	&&\!\!\!\!\cI_{(m,n;p,q)}\\
	=&&\!\!\!\!\langle \textsf{A}_m\textsf{B}_p \rangle+\langle \textsf{A}_m\textsf{B}_q \rangle+\langle \textsf{A}_n\textsf{B}_p \rangle-\langle \textsf{A}_n\textsf{B}_q \rangle\leq 2,\nonumber
\end{eqnarray}
where the indices $m,n,p,q$ label the inputs. The correlators are defined as 
\begin{eqnarray}
	\label{eq:correlator}
\langle \textsf{A}_x\textsf{B}_y \rangle=\sum_{a,b}ab\cdot p(a,b|x,y).
\end{eqnarray}
Generally, the quantum correlations $p(a,b|x,y)$ may be derived from several different combinations of states and local measurements. To self test the target state $\ket{\ps'}$, we must find correlations that are produced uniquely by $\ket{\ps'}$ up to a certain equivalence class. Hence, we can certify the state $\ket{\ps'}$ up to equivalence from the quantum correlations only. In this paper, the equivalence class is characterized by a local unitary operator $U$. By the CHSH inequalities, the statement of self testing for pure state is given as follows.
\begin{definition}
	\label{def:ST}	
A state $\ket{\ps'}\in\cH^{A'}\otimes\cH^{B'}$ can be self-tested by the correlations $p(a,b|x,y)$ if for any quantum realization of them, there exists a local unitary $U$  	such that 
the state
\begin{eqnarray}
	[\ket{\xi_0}^{AB}\otimes\ket{00}^{A''B''}+\ket{\xi_1}^{AB}\otimes\ket{11}^{A''B''}]\otimes\ket{\ps'}^{A'B'}
\end{eqnarray}
 can be extracted when the local unitary is applied to the unknown state $\ket{\ps}\in\cH^{A}\otimes\cH^{B}$ of the system together with local auxiliary states $\ket{00}\in[\cH^{A''}\otimes \cH^{A'}]\otimes[\cH^{B''}\otimes\cH^{B'}]$, where  $\ket{\xi_j}$ are some unknown subnormalized junk states such that $\norm{\ket{\xi_0}}+\norm{\ket{\xi_1}}=1$.
\end{definition}
 The key for self testing is to construct the local unitary $U$ so that the statements following Definition \ref{def:ST} can be proved. Once the CHSH inequality in Eq. (\ref{eq:cI}) is maximally violated, i.e. the LHS of Eq. (\ref{eq:cI}) reaches $2\sqrt{2}$, then the existence of the desired local unitary can be proved. 
 
In a recent work, it has been proved that any multipartite entangled states can be self tested  \cite{ivan2023quantum}. In this paper, we only need the self testing for bipartite maximally entangled state and certify the presence of tensor products of Pauli observables on maximally entangled states, which is shown in Refs. \cite{bowles2018pra, bowles2018prl}. In detail, Alice has a choice of three measurement operators $\textsf{A}_x$ corresponding to the input $x=1,2,3$. Bob has a choice of six measurements $\textsf{B}_y$ corresponding input $y=1,2,...,6$. The following 3-CHSH inequality  is utilized
\begin{eqnarray}
	\label{eq:3CHSH}
\b_{CHSH}=\cI_{(1,2;1,2)}+\cI_{(1,3;4,3)}+\cI_{(2,3;6,5)}\leq6,
\end{eqnarray}
where $\cI_{(m,n;p,q)}$ is defined as Eq. (\ref{eq:cI}). Each of Alice's inputs appears in two of the three CHSH inequalities. For convenience of the statement, we denote $I$, $\s_z$, $\s_x$, $\s_y$ as $\s_0$, $\s_1$, $\s_2$, $\s_3$, respectively. A fact about this self-testing process has been proved, that is,  the maximal violation of the 3-CHSH inequality in Eq.  (\ref{eq:3CHSH}) can be achieved by the following states and observables,
  \begin{eqnarray}
  	&&
  	\label{eq:psphi+} \ket{\ps'}=\ket{\Phi^0}=\frac{1}{\sqrt{2}}(\ket{00}+\ket{11}),\\
  	\label{eq:XAYAZA}
  	&&\textsf{A}_x=\s_x, \quad x=1,2,3,\\
  	\label{eq:DBjk}
  	&&\textsf{B}_{y}=\frac{\s_j\pm\s_k}{\sqrt{2}}, 
  	 \quad y=1,2,...,6,
  \end{eqnarray} 
  for $(j,k)\in \{(1,2),(1,3), (2,3)\}$. 
  The reverse statement, that the maximal violation $6\sqrt{2}$ can only be achieved by the measurements applied on $\ket{\Phi^0}$,  representing the first self-testing statement \cite{yao2004self}. 
  
 Using the techniques above, we can certify the presence of tensor products of Pauli observables on maximally entangled states \cite{bowles2018pra, bowles2018prl}.
\begin{lemma}
	\label{le:ST observables}
	Let  Alice and Bob share the state $\ket{\ps}\in\cH^A\otimes \cH^B$ and binary observables $\textsf{A}_x$ in (\ref{eq:XAYAZA}) and $B_y$ in (\ref{eq:DBjk}), respectively. If one observes the maximal violation of CHSH inequality, i.e. $\b_{CHSH}=6\sqrt{2}$, then there exist local auxiliary states $\ket{00}\in[\cH^{A''}\otimes \cH^{A'}]\otimes[\cH^{B''}\otimes \cH^{B'}]$ and a local unitary $U=U^{AA'A''}\otimes U^{BB'B''}$ such that 
	\begin{eqnarray}
&&\!\!\!\!\!\!U[\textsf{A}_1\ket{\ps}^{AB}\otimes\ket{00}]=\ket{\xi}=\otimes\s_1^{A'}\ket{\Phi^0}^{A'B'},
		\\
		&&\!\!\!\!\!\!U[\textsf{A}_2\ket{\ps}^{AB}\otimes\ket{00}]=\ket{\xi}\otimes\s_2^{A'}\ket{\Phi^0}^{A'B'},
		\\
		\label{eq:selftestY}
		&&\!\!\!\!\!\!U[\textsf{A}_3\ket{\ps}^{AB}\otimes\ket{00}]=\s_1^{A''}\ket{\xi}\otimes\s_3^{A'}\ket{\Phi^0}^{A'B'},
	\end{eqnarray}	
	where the state $\ket{\xi}\in[\cH^{A}\otimes \cH^{A''}]\otimes[\cH^{B}\otimes \cH^{B''}]$ takes the form
	\begin{eqnarray}
		\ket{\xi}=\ket{\xi_0}^{AB}\otimes\ket{00}^{A''B''}+\ket{\xi_1}^{AB}\otimes\ket{11}^{A''B''}
	\end{eqnarray}
with the unknown subnormalized junk states $\ket{\xi_j}$, for $j=0,1$.
\end{lemma}
 The proof of Lemma \ref{le:ST observables}  are provided in Refs. \cite{bowles2018pra}. 
Note that we can only certify the Pauli observables up to conjugation in the computational basis. In Eq. (\ref{eq:selftestY}), one can see that the measurement $\textsf{A}_3$ can be treated as  first measuring an observable $\s_1$ on qubit $A'' $ of the state $\ket{\xi}$,  whose  outcome decides either $\s_3$ or $-\s_3$ is performed on the state $\ket{\Phi^0}$.  That is, the action of $\textsc{A}_3$ is to perform $\s_3$ and $-\s_3$ with the probability $\braket{\xi_0}{\xi_0}$ and $\braket{\xi_1}{\xi_1}$, respectively.
 
Self-testing entangled measurements is a key   elements for quantum networks. In an entanglement swapping scenario, the certification of Bell measurements  is given in Refs. \cite{Renou2018self}. We restate the main results for the certification of reference Bell measurement given by $\{\proj{\Phi^b}\}_{b=0,1,2,3}$ with
\begin{eqnarray}
\ket{\Phi^0}=(\ket{00}+\ket{11})/\sqrt{2},\;	\ket{\Phi^1}=(\ket{00}-\ket{11})/\sqrt{2},\\
\ket{\Phi^2}=(\ket{01}+\ket{10})/\sqrt{2},\;	\ket{\Phi^3}=(\ket{01}-\ket{10})/\sqrt{2}.
\end{eqnarray}
 Three parties Alice, Bob and Charlie are employed in the certification. Alice and Charlie choose the local measurements $\textsf{A}_j=\s_j$  and $\textsf{C}_j=(\s_1\pm\s_2)/\sqrt{2}$ with $j=1,2$. If the statistics conditioned on the result $b$ lead to the maximal violation $\g_{CHSH_b}=2\sqrt{2}$, where
\begin{eqnarray}
\label{def:gchsh0}
&&\!\!\!\!\!\!\!\!\!\g_{CHSH_0}=\langle \textsf{A}_1\textsf{C}_1 \rangle+\langle \textsf{A}_1\textsf{C}_2 \rangle+\langle \textsf{A}_2\textsf{C}_1 \rangle-\langle \textsf{A}_2\textsf{C}_2 \rangle,\\
&&\!\!\!\!\!\!\!\!\!\g_{CHSH_1}=\langle \textsf{A}_1\textsf{C}_1 \rangle+\langle \textsf{A}_1\textsf{C}_2 \rangle-\langle \textsf{A}_2\textsf{C}_1 \rangle+\langle \textsf{A}_2\textsf{C}_2 \rangle,\\
\label{def:gchsh23}
&&\!\!\!\!\!\!\!\!\!\g_{CHSH_2}:=-\g_{CHSH_1},\quad\g_{CHSH_3}:=-\g_{CHSH_0}.
\end{eqnarray}   
then it implies that Bob performs the Bell state measurement as the reference one, up to some local isometries.
\begin{lemma}
	\label{le:certifyBSM}
Let the initial state shared by Alice, Bob and Charlie be of the form $\tau^{AB_1B_2C}=\tau^{AB_1}\otimes\tau^{B_2C}$, and let $\cB:=\{M_b^{B_1B_2}\}_{b=0,1,2,3}$ be a two-outcome measurement acting on $\cH^{B_1}\otimes \cH^{B_2}$. If there are measurements for Alice and Charlie such that the resulting correlations exhibit the maximal violation of the CHSH inequality $\g_{CHSH_b}$, then there exists completely positive and unital maps $\L_{B_1}:\cL(\cH^{B_1})\rightarrow\cL(\cH^{A'})$, $\L_{B_2}:\cL(\cH^{B_2})\rightarrow\cL(\cH^{C'})$ for $\abs{A'}=\abs{C'}=2$ such that
\begin{eqnarray}
(\L_{B_1}\otimes\L_{B_2})(M_b^{B_1B_2})=\proj{\Phi^b}^{A'C'}.
\end{eqnarray}
\end{lemma}
 The above results formalize the notion that the real measurement is in some sense equivalent to the reference one.  Note that the dual map $\L_{B_1}^\dagger\otimes\L_{B_2}^\dg: \cL(\cH^{A'})\otimes\cL(\cH^{C'})\rightarrow\cL(\cH^{B_1})\otimes\cL(\cH^{B_2}) $ is completely positive and trace preserving. Given an unknown state $\mu$ acting on $\cH^{A'C'}$, we would like to obtain the statistics produced under the ideal measurement $\proj{\Phi^b}^{A'C'}$. It suffices to apply this dual map to $\mu$ and perform the real measurement $M_b^{B_1B_2}$. In fact, the probability of observing the outcome $k$ is given by
 \begin{eqnarray}
 p(k)= &&\!\!\!\!\!\!\tr[\L_{B_1}^\dagger\otimes\L_{B_2}^\dg(\mu)M_b^{B_1B_2}]\\
 =&&\!\!\!\!\!\!\tr[\mu \proj{\Phi^b}^{A'C'}].
 \end{eqnarray}

\section{Device-independent discrimination of quantum states}
\label{sec:discrimination}
In this section we show the DI multiple-copy state discrimination protocol for quantum states using the self testing results.   First we show the discrimination of qubit pure states in the black-box scenario. On the basis of that, we establish the network for discrimination of $n$-qubit pure states. Without loss of generality, two  arbitrary states that are not conjugate with each other can be discriminated  in a fully DI scenario, while others can be discriminated measurement device independently. 

\subsection{DI  discrimination of single-qubit states}
\label{sec:1qubit discrimination}
\begin{figure}[htp]
	\center{\includegraphics[width=8.5cm]  {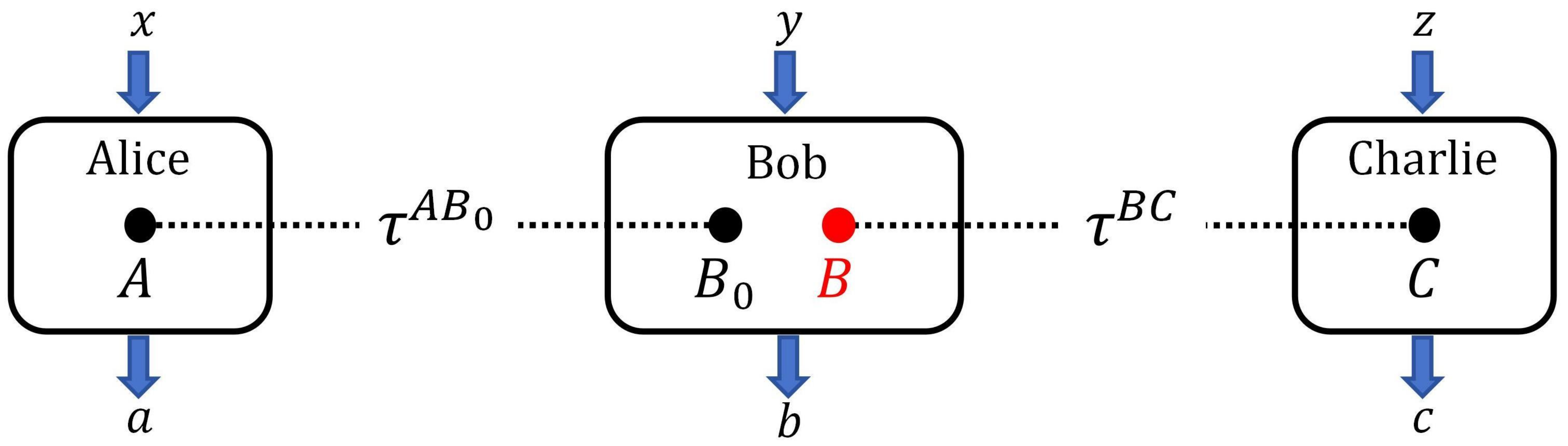}}
	\caption{Quantum network for DI multi-copy quantum state discrimination of single-qubit states. The quantum correlations $p_1(a,b,c|x,y,z)$ and $p_2(a,b|x,\diamond)$ are generated by collecting the statistics for inputs $x=1,2,3$, $y=1,2,...,6$, $z=1,2$ and outputs $a,b,c=\pm 1$.} 
	\label{fig:network1}
\end{figure}

We present the DI discrimination protocol of qubit states, which works in the following scenario. One of the single-qubit pure state $\r^B$ is chosen from the ensemble $\{q_j, \ps_j^B\}_{j=1}^Q$. Then it is sent to the main party of this protocol, named Bob, who aims to discriminate this state together with two auxiliary parties. However, the devices available to Bob are all untrusted, which implies that the outcomes produced by the devices may be deceptive.   To tackle such a problem, the auxiliary parties Alice and Charlie are added in this protocol. They perform local measurements and produce the statistics. Since all parties do not communicate with each other, a referee of this game will collect these statistics and generate the correlations. Based on the correlations, the experiment devices will be certified and the result of quantum state discrimination will be obtained. The DI quantum state discrimination can be realized when the expected correlations are observed. Otherwise, nothing will be obtained due to the uncertainty of the security for devices.  It is realized by the network featuring two bipartite states, shown in FIG. \ref{fig:network1}.

In this scenario, the parties Alice, Bob and Charlie simply model the laboratories as a black box which takes an input and returns an output. All they know about the system is that the quantum state $\r^B$ is from a known ensemble $\{q_j, \r_j^{B}\}_{j=1}^Q$.  In the DI scenario, two assumptions are necessarily required:
\begin{enumerate}
	\item Measurements and states are corrected described by the network in FIG. \ref{fig:network1}. They follow the nature of quantum mechanics.
	\item The experiment setup is  independent and identically distributed, i.e.   the distributed quantum states and strategies are the same for each round of experiments.
\end{enumerate} 

Next we show how the network is used for the discrimination of single-qubit states. Specifically, Bob shares the auxiliary states $\tau^{AB_0}$ and $\tau^{BC}$ with Alice and Charlie, respectively.  In order to  certify the measurements for systems $A$ and $B_0B$, each party of this protocol performs the local measurements  and obtains the device certification correlations
\begin{eqnarray}
	\label{eq:p(ab)1}
	&&p_1(a,b,c|x,y,z)\\
	=&&\tr[M_{a|x}^A\otimes M_{b|y}^{B_0B}\otimes M_{c|z}^{C} \tau^{AB_0} \otimes \tau^{BC} ]
\end{eqnarray}
where the local measurements $M_{j|k}$ denote the projector corresponding to the measurement $k$ with outcome $j$. Once the certification correlations lead to the maximal violation of CHSH inequalities, the state for system $B$ is replaced with the target state $\r^B$. Since all devices remain the same in each round of experiments,  the correlations for quantum state discrimination are given by
\begin{eqnarray}
	\label{eq:p(ab)12}
p_2 (a,b|x,\diamond)=\tr[M_{a|x}^A\otimes M_{b|\diamond}^{B_0B}\tau^{AB_0} \otimes \r^{B}].
\end{eqnarray} 
Note that all measurements and states in this network are untrusted, while the  certification of devices and results for quantum state discrimination are trustworthy. It is realized based on the reliable correlations given in Eqs. (\ref{eq:p(ab)1}) and (\ref{eq:p(ab)12}). To obtain the correlations, multi copies of states $\r^{B}$ are required, and the success rate of discrimination approaches one when enough copies are provided. Note that the discrimination of non-orthogonal states is included. Hence the requirement for "many copies of target states" is natural for them. 

We show the framework of DI quantum state discrimination protocol for single-qubit states. More details are provided later.
\begin{center}
	\begin{tcolorbox}[colback=gray!0,		
		colframe=black,					
		width=8.5cm,							
		arc=1mm, auto outer arc,            
		boxrule=1pt]                        
		\textcolor{black}                   
		{{\textbf{Protocol} The DI quantum state discrimination protocol for arbitrary qubit states can be realized by the network in FIG. \ref{fig:network1}.
		Three steps for this protocol are listed as follows:
		\begin{enumerate}
\item Implementation of measurements and collection of correlations.   \\
All parties perform local measurements and obtain the correlations $p_1(a,b,c|x,y,z)$ and $p_2(a,b|x,\diamond)$, where $x=1,2,3$, $y=1,2,...,6,\diamond$, $z=1,2$ and $a,b,c=\pm 1$.
\item Certification of auxiliary states and measurements via self testing.\\
 First, the  correlations $p_1(a,b,c|x,y,z)$ with $y=1,2,...,6$ certify that auxiliary state $\tau^{AB_0}$ contains the maximally entangled states, and Alice performs the Pauli observables on the local systems. Second, the correlations $p_1(a,b,c|x,\diamond,z)$ certify that the Bob's two-qubit measurement $M^{B_0B}_{b|\diamond}$ is equivalent to the reference Bell measurement $\proj{\Phi^b}$.
\item DI quantum state discrimination. \\
The states from a known ensemble $\{q_j, \r_j^{B}\}_{j=1}^Q$  are distinguished by a certain correspondence between the target state $\r^B$ and different correlations $P_2(a|x)$, which is derived by $p_2(a,b|x,\diamond)$.
\end{enumerate}
		}}
	\end{tcolorbox}
\end{center}
Next we show the details for each step of this protocol. 

Step 1:  We show the implementation of measurements and collection of correlations. In this protocol, Alice has a choice of three measurements corresponding to inputs $x=1,2,3$. Bob and Charlie have the choice of inputs $y=1,2,...,6,\diamond$ and $z=1,2$, respectively. Based on the facts provided in Sec. \ref{sec:pre,ST}, the measurements corresponding to different inputs are chosen as  
\begin{eqnarray}
	\label{eq:AX}
&&\textsf{A}_x=\s_x, \quad x=1,2,3,\\
&&\textsf{B}_{y}=\frac{\s_j\pm\s_k}{\sqrt{2}}, 
\quad j,k=1,2,3,\; y=1,2,...,6,\\
\label{eq:C12}
&&\textsf{C}_1=\frac{\s_1+\s_2}{\sqrt{2}},\; \textsf{C}_2=\frac{\s_1-\s_2}{\sqrt{2}}.
\end{eqnarray}
Bob manipulates two qubits $B_0$ and $B$. When Bob receives the inputs $y=1,2,...,6$, the single-qubit measurement $\textsf{B}_y$ is only performed on qubit $B_0$. The two-qubit measurement operators for inputs
 $ y=\diamond$ and outcomes $b$ are defined as $M_{b|\diamond}^{B_0B}=\proj{\Phi^b}$, for $b=0,1,2,3$.   Besides, no restrictions are made on the auxiliary states and they can be chosen to be $\tau^{AB_0}=\tau^{BC}=\proj{\Phi^0}$.

Step 2: We  certify the auxiliary states and measurements based on the self testing results in Sec. \ref{sec:pre,ST}.
First, the auxiliary state $\tau^{AB_0}$ and Alice's local measurements are certified using the correlations $p_1(a,b,c|x,y,z)$ with $y=1,2,...,6$.   Recall that the two-qubit measurement $M_{b|y}^{B_0B}=N_{b|y}^{B_0}\otimes I^B$. From Eq. (\ref{eq:p(ab)1}), the correlations with $y=1,2,...,6$ are given as
\begin{eqnarray}
&&	p_1(a,b,c|x,y,z)\\
	=&&\tr[M_{a|x}^A\otimes N_{b|y}^{B_0}\otimes I^B \otimes M^C_{c|z} \tau^{AB_0} \otimes \tau^{BC} ]\nonumber\\
	=&&\tr[M_{a|x}^A\otimes N_{b|y}^{B_0} \tau^{AB_0}]\nonumber\\
	:=&&p_1(a,b|x,y).\nonumber
\end{eqnarray}
It implies that the correlations $p_1(a,b,c|x,y,z)$ can be used directly for the certification of systems $A$ and $B_0$. From Lemma \ref{le:ST observables}, if the statistics of $p_1(a,b,c|x,y,z)$ lead to maximal violation of the 3-CHSH inequalities
\begin{eqnarray}
	\label{ieq:bchsh}
\b_{CHSH}=\cI_{(1,2;1,2)}+\cI_{(1,3;4,3)}+\cI_{(2,3;6,5)}\leq6,
\end{eqnarray}
where
\begin{eqnarray}
	\cI_{(m,n;p,q)}=\langle \textsf{A}_m\textsf{B}_p \rangle+\langle \textsf{A}_m\textsf{B}_q \rangle+\langle \textsf{A}_n\textsf{B}_p \rangle-\langle \textsf{A}_n\textsf{B}_q \rangle
\end{eqnarray}
and
\begin{eqnarray}
	\langle \textsf{A}_x\textsf{B}_y \rangle=\sum_{a,b}ab\cdot p_1(a,b|x,y).
\end{eqnarray}
then the auxiliary states $\tau^{AB_0}$ contain the maximally entangled state $\proj{\Phi^0}$, and Pauli measurements $\s_1,\s_2,\s_3$ for system $A$ can be certified up to local conjugation. 

Second, we show the certification of the measurement $M^{B_0B}_{b|\diamond}$. The  CHSH inequalities given in Eqs. (\ref{def:gchsh0})-(\ref{def:gchsh23}) with $\langle \textsf{A}_x\textsf{C}_z \rangle=\sum_{a,b,c}abc\cdotp p(a,b,c|x,\diamond,z)$ are used here.
 From Lemma \ref{le:certifyBSM}, the maximally violation of $\g_{CHSH_b}$ implies that Bob performs the Bell state measurement as the reference measurement $\proj{\Phi^b}$, up to some local isometries.

Step 3: We introduce the quantum state discrimination based on the correlations $p_2(a,b|x,\diamond)$ given in Eq. (\ref{eq:p(ab)12}). We establish a certain correspondence between states and  different correlations such that any bipartite pure state from the ensemble  $\{q_j, \ps_j^{AB}\}_{j}$ can be depicted by unique correlations. On the basis of that, we can distinguish different states from the observed statistics collected from Step 1. 
When the maximal violation of CHSH inequalities in Eqs. (\ref{def:gchsh0})-(\ref{def:gchsh23}) and (\ref{ieq:bchsh})  is observed, the uncharacterized measurements $M_{a|x}$, $M_{b|\diamond}$ and state $\tau^{AB_0}$ in  Eq. (\ref{eq:p(ab)12}) are certified as we have mentioned in Step 2.
Using Lemma \ref{le:ST observables} and \ref{le:certifyBSM}, we obtain the correlation as follows
\begin{eqnarray}
	\label{eq:p(a,b)}
	&&p_2(a,b|x,\diamond)\\
	=&&\tr[\pi_{a|x}\otimes \proj{\Phi^b}
	(\proj{\Phi^0}\otimes \r^{B})]\\
	=&&\frac{1}{2}\tr[ \proj{\Phi^b}
	\pi_{a|x}^{T} \r^{B}]\\
	=&&\frac{1}{4}\tr[
	U_b^\dg\pi_{a|x}U_b \r^{B} ],
\end{eqnarray}
where the unitaries $U_0=\bbI, U_1=\s_1, U_2=\s_2, U_3=\s_1\s_2$, and the measurement projectors  $\pi_{j|k}=\frac{1}{2}[I+j\s_k]$ with $j=\pm 1$ and $k=1,2,3$. The second and third equations are derived by 
\begin{eqnarray}
	\label{eq:trphi}
\tr_M[\proj{\Phi^0}^{MN}\pi_{i|j}^M\otimes\bbI^N]=\frac{1}{2}(\pi_{i|j}^N)^T.	
\end{eqnarray} 
We will show that the correlations
\begin{eqnarray}
	\label{eq:correP2(a|x)}
P_2^{\r}(a|x):=\tr[\pi_{a|x}\r]=4p_2(a,0|x,\diamond)
\end{eqnarray}  
are sufficient to distinguish the quantum states. That is, we can always establish the correspondence between states and correlations $P_2(a|x)$ such that the correlations differ only when the states differ. 

We show the derivation of $P_2(a|x)$ experimentally. It can be obtained that $U_0\pi_{a|x}U_0=\pi_{a|x}$ and 
\begin{eqnarray}
	\label{eq:UbpiUb}
U_b^\dg\pi_{a|x}U_b
=\begin{cases}
	\pi_{a|x},\;x=b,\\
	\pi_{-a|x},\;x\neq b,
\end{cases}
\end{eqnarray}
for $b=1,2,3$.
The correlations $p_2(a,b|x,\diamond)$ satisfy that
\begin{eqnarray}
	\label{eq:p2(a01d)}
p_2(a,b|x,\diamond)=\begin{cases}
p_2(a,0|x,\diamond),\;x=b,\\
p_2(-a,0|x,\diamond),\;x\neq b.
\end{cases}
\end{eqnarray}
 Hence, 
\begin{eqnarray}
\label{eq:P2(a1)}
P_2(a|1)=&& \!\!\!\!\!\! p_2(a,0|1,\diamond)+p_2(a,1|1,\diamond)\\
+&& \!\!\!\!\!\!p_2(-a,2|1,\diamond)+p_2(-a,3|1,\diamond),\nonumber\\
P_2(a|2)=&&\!\!\!\!\!\! p_2(a,0|2,\diamond)+p_2(-a,1|2,\diamond)\\+&& \!\!\!\!\!\!p_2(a,2|2,\diamond)+p_2(-a,3|2,\diamond),\nonumber\\
\label{eq:P2(a3)}
P_2(a|3)=&& \!\!\!\!\!\! p_2(a,0|3,\diamond)+p_2(-a,1|3,\diamond)\\
+&&\!\!\!\!\!\! p_2(-a,2|3,\diamond)+p_2(a,3|3,\diamond).\nonumber
\end{eqnarray}
It shows that the referee can classify the inputs and outputs in each line of Eqs. (\ref{eq:P2(a1)})-(\ref{eq:P2(a3)}) into  the same category and thus obtains the corresponding correlations $P_2(a|x)$. For example, when the referee aims to obtain the correlation $P_2(1|1)$, he will focus on the number of occurrences corresponding to input $(1,\diamond)$ and outputs $(1,0),(1,1),(-1,2),(-1,3)$, dividing which  by the total rounds of experiment will obtain the desired correlation. We use the correlations $P_2(a|x)$ instead of $p_2(a,0|x,\diamond)$ here so that the statistics can be fully utilized.

Next we show the DI discrimination of states using the correlations $P_2(a|x)$. The distinguishability between the target states under all possible measurements is conventionally measured by trace norm \cite{bae2017Quantum}. For the set  of POVMs $A=\{A_x\}_{x=1,2,...}$, the  trace norm of $X$ is
$\norm{X}_1:=\max_A \abs{\tr[AX]}= \max_{A_x\geq 0, \sum_xA_x=\bbI}\sum_x\abs{\tr(A_xX)}$. 
On the basis of that, the distinguishability of states under specific measurement strategy can also be characterized.  Restricting the possible measurements on a set of informationally complete POVMs, denoted by $\cM$, a norm induced by the measurements can be constructed as \cite{Matthews2009distinguishability}
\begin{eqnarray}
	\label{def:normX_M}
	\norm{X}_{\cM}:=&&\!\!\!\!\!\!\max_\cM\abs{\tr[\cM X]}\\
	=&&\!\!\!\!\!\! \max_{\cM=\{{\cM}_a\}_{a=1}^n}\sum_a\abs{\tr({\cM}_aX)}.
\end{eqnarray}   
Obviously we have  $\norm{X}_{\cM}\leq\norm{X}_1$, and the trace norm can be recovered by setting $\cM$ as the most general measurement strategy.
The projectors $\pi_{j|k}$ are on the plus and minus eigenspace of Pauli matrices. Three sets of POVMs form the informationally complete set
\begin{eqnarray}
	\label{def:tomogrphy set}
	\cN=\Big\{&&\!\!\!\!\!\!
	\cM^{(1)}=\{\pi_{1|1},\pi_{-1|1}\}, \;       \cM^{(2)}=\{\pi_{1|2},\\
	&&\!\!\!\!\!\! \pi_{-1|2}\},\; \cM^{(3)}=\{\pi_{1|3},\pi_{-1|3}\}\Big\}. \nonumber
\end{eqnarray}
Choosing $\cM=\cN$,  it holds that $\norm{\cdot}_{\cN}$ constitutes a norm.

We show the process of single-qubit  state discrimination in detail. 
Suppose Bob receives a state for system $B$ that is known to be from the ensemble consisting $Q$ pure states, 
\begin{eqnarray}
\label{eq:r1r2}
\{p_j, \r_j=\proj{\a_j}\}_{j=1}^Q
\end{eqnarray}
		 where
$\ket{\a_j}=\cos\og_j\ket{0}+e^{i\t_j}\sin\og_j\ket{1}$
with $\og_j\in [0,\pi/2]$ and $\t_j\in[0,2\pi)$. 
We consider the discrimination of these pure states. It suffices to analyze two different states denoted by $\r_1$ and $\r_2$ in this ensemble. 
 The bias of probability is given by
\begin{eqnarray}
\Delta_x=&&\!\!\!\!\!\!\abs{\tr[\cM^{(x)}(\r_1-\r_2)]}\\
=&&\!\!\!\!\!\!\abs{P_2^{\r_1}(a|x)-P_2^{\r_2}(a|x)}.
\end{eqnarray}
Using (\ref{def:normX_M}) and (\ref{def:tomogrphy set}),  the distance between the target states is derived by 
\begin{eqnarray}
	&&\!\!\!\!\!\! d_{\cN}(\r_1,\r_2)
	:=\frac{1}{2}\norm{\r_1-\r_2}_{\cN}=\frac{1}{2}\max_x\{\Delta_x\}
\end{eqnarray} 
where 
\begin{eqnarray}
	\label{eq:cosog1}
	\Delta_1=&& \!\!\!\!\!\!2\abs{\cos^2\og_1-\cos^2\og_2},\\
	\Delta_2=&& \!\!\!\!\!\!\abs{\mbox{Re}(e^{-i\t_1}\sin2\og_1 -e^{-i\t_2}\sin2\og_2)},\\
	\label{eq:Imsin2og} \Delta_3=&& \!\!\!\!\!\!\abs{\mbox{Im}(e^{-i\t_1}\sin2\og_1 -e^{-i\t_2}\sin2\og_2)}. 
\end{eqnarray} 
Obviously $\Delta_x$ correspond to the Pauli observable $\s_x$, respectively.
Two states can be distinguished by determined Pauli observables when their distance $d_{\cN}>0$, and the distinguishability between them increases with $d_{\cN}$. The informational
 completeness of the set $\cN$ naturally allows the following fact. 
\begin{lemma}
	\label{le:single qubit}
	Two different single-qubit pure states $\r_1$ and $\r_2$ can be discriminated by POVMs from the set $\cN$ in Eq.  (\ref{def:tomogrphy set}).   That is, there exist $x\in\{1,2,3\}$ such that
	\begin{eqnarray}
d_{\cN}(\r_1,\r_2)=\Delta_x/2>0,
	\end{eqnarray}
where $a=1$ or $-1$ and the correlation $P_2(a|x)$ is defined in (\ref{eq:correP2(a|x)}). 
\end{lemma}
From Lemma \ref{le:single qubit}, we can always discriminate the qubit states by three determined Pauli observables. Lemma \ref{le:ST observables} shows that the Pauli observables are certified up to  conjugation. Hence, the projectors are undetermined and we cannot use Lemma \ref{le:single qubit} directly. 
To be specific,  the Pauli observables $\s_1$ and $\s_2$ can be self tested exactly, while it is uncertain whether $\s_3$ or $-\s_3$ is performed on the maximally entangled state. So it fails to determine some of the states when the correlations $P_2^{\r}(a|3)$ is necessarily required. From Eqs. (\ref{eq:cosog1})-(\ref{eq:Imsin2og}), when the coefficients of states show $\og_1=\og_2$ and $\t_1=2\pi-\t_2$, the correlations $P_2^\r(a|3)$ are necessarily required.  So the target states can only be characterized up to configuration.  For example, when the target state  $\r=\proj{R}$ is measured by  observable $\s_3$, the correlations show $P_2^{\r}(+1|3)=1/4$ and $P_2^{\r}(-1|3)=0$, while it shows that $P_2^{\r}(+1|3)=0$ and $P_2^{\r}(-1|3)=1/4$ when the observable $-\s_3$ is performed.

The above analysis shows that two states conjugate with each other cannot be discriminated due to the conjugation of $\s_3$. Since the self-test of  other measurements also holds up to configuration, the discrimination of these states can not be realized device independently. Next we show that it can be realized in a measurement device independent (MDI) manner. That is, Bob prepares a single copy of the trusted state $\eta=\proj{R}$ and add it into the sequence of target states.  With the input $x=3$, if the output for $\eta$ is +1 (resp. $-1$), the referee confirms that $\textsf{A}_3$ in (\ref{eq:selftestY}) acts as $\s_3$ (resp. $-\s_3$). So these states can be discriminated successfully.

\begin{figure*}[htp]
	\center{\includegraphics[width=14cm]  {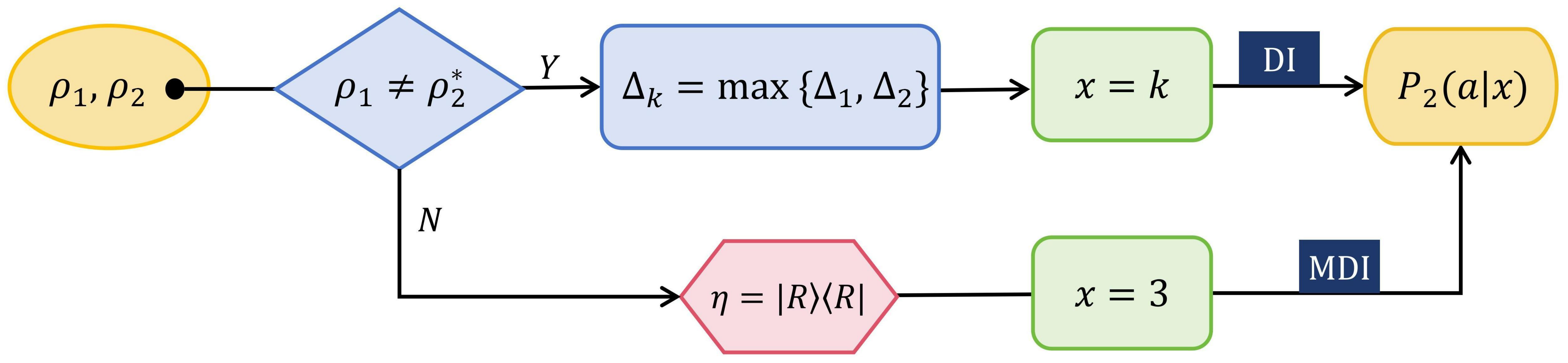}}
	\caption{ Diagram showing  the best choice of correlations for the discrimination of qubit pure states $\r_1,\r_2$ defined in Eq. (\ref{eq:r1r2}). For the states that are not conjugate with each other, the DI discrimination is realized by correlation $P_2(a|k)$, where $k$ is the index satisfying  $\Delta_k=\max\{\Delta_1,\Delta_2\}$. The states conjugate with each other can be discriminated by correlation $P_2(a|3)$ in a MDI manner, as the trusted state $\eta$ is used here.
	By such a choice, each kind of correlations differ for the states $\r_1$ and $\r_2$, and the correspondence is given with certainty. } 
	\label{fig:diagram1qubit}
\end{figure*}

To sum up, DI discrimination can be realized for two quantum sates that are not conjugate with each other, and MDI discrimination can be realized for other states.  This process and the best choice of the correlations  $P_2^\r(a|x)$ are shown in FIG. \ref{fig:diagram1qubit}.  It implies that the referee can only choose the input $p\in\{1,2,3\}$ and observe the statistics generated by the measurement operator $\s_p$ for the  discrimination of states. Hence fewer copies of target states are required than tomography.
The discrimination of multiple states can be realized in a similar way.

\subsection{DI discrimination of $N$-qubit states}
\begin{figure}[htp]
	\center{\includegraphics[width=8.4cm]  {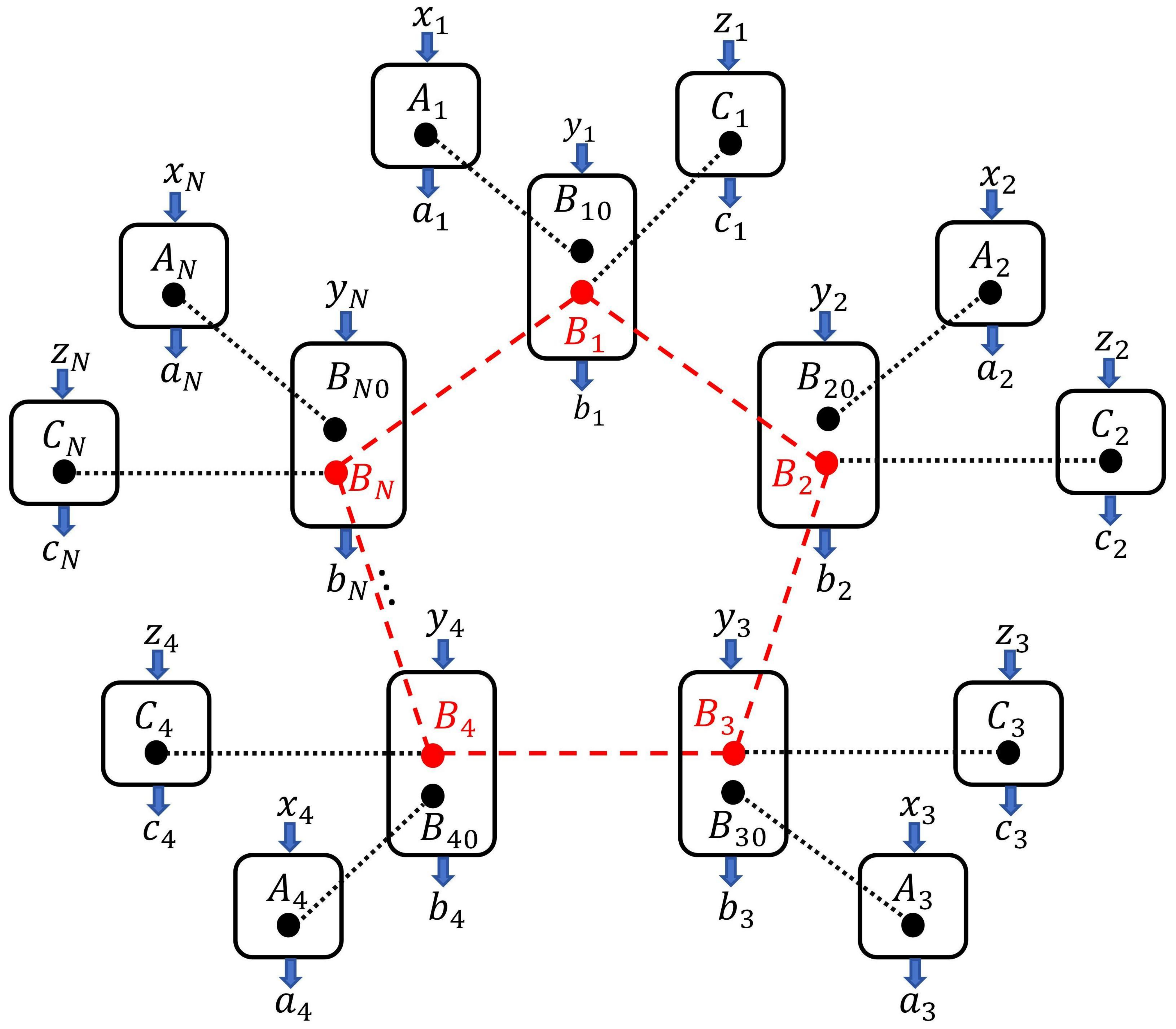}}
	\caption{Quantum network for DI multi-copy  state discrimination of $N$-qubit states. The  correlations $p_1(\texttt{a,b,c|x,y,z})$ and $p_2(\texttt{a,b|x},\diamond)$ are generated by collecting the statistics for inputs and outputs of this network.} 
	\label{fig:networkNqubit}
\end{figure}

The DI discrimination of multi-qubit states can be realized by extending the protocol for single-qubit state in Sec. \ref{sec:1qubit discrimination}. Suppose the main parties $B_1,B_2,...,B_N$ receive the $N$-qubit pure state $\r^{B_1B_2...B_N}$, which is from the ensemble 
\begin{eqnarray}
	\{q_j, \ph_j\}_{j=1}^Q	
\end{eqnarray}
with
\begin{eqnarray}
\label{def:phj}
\ph_j=\frac{1}{2^N}\sum_{j_1j_2...j_N}S^{(j)}_{j_1j_2...j_N}\s_{j_1}\otimes\s_{j_2}...\otimes\s_{j_N}.
\end{eqnarray}
In order to figure out which state the main parties are sharing, the auxiliary parties $A_k$ and $C_k$ are introduced, who share the auxiliary states $\tau^{A_kB_{k0}}$ and $\tau^{B_kC_k}$ with the main party $A_k$, for $k=1,2,...,N$.    On the inputs  $\texttt{x}=(x_1,x_2,...,x_N)$, $\texttt{y}=(y_1,y_2,...,y_N)$ and $\texttt{z}=(z_1,z_2,...,z_N)$, the parties $\{A_j\}_{j=1}^N, \{B_j\}_{j=1}^N, \{C_j\}_{j=1}^N$ perform measurements defined in Eqs. (\ref{eq:AX})-(\ref{eq:C12}) and return the outputs  $\texttt{a}=(a_1,a_2,...,a_N)$,  $\texttt{b}=(b_1,b_2,...,b_N)$ and  $\texttt{c}=(c_1,c_2,...,c_N)$ respectively, where $x_j=1,2,3$, $y_j=1,2,...,6,\diamond$, $z_j=1,2$, and $a_j,b_j,c_j=\pm 1$. The network used to realize the $N$-qubit state discrimination is presented in FIG. \ref{fig:networkNqubit}.

As the discrimination of single-qubit state, 
 all parties perform the measurements to generate the correlations $p_1(\texttt{a,b,c|x,y,z})$ for device certification firstly. They are given by
 \begin{eqnarray}
&&\!\!\!\!\!\!p_1(\texttt{a,b,c|x,y,z})\\
=&&\!\!\!\!\!\!\tr\left[\bigotimes_{k=1}^N(M_{a_k|x_k}^{A_k}\otimes M_{b_k|y_k}^{B_{k0}B_k}\otimes M_{c_k|z_k}^{C_k}\tau^{A_kB_{k0}}\otimes\tau^{B_kC_k})\right]\nonumber\\
=&&\!\!\!\!\!\! \prod_{k=1}^N p_1(a_k,b_k,c_k|x_k,y_k,c_k)
:=\prod_{k=1}^N p_1(a_k,b_k|x_k,y_k).\nonumber
 \end{eqnarray}

Then they replace the state for systems $B_k$ by the target state $\r^{B_1B_2...B_N}$ to obtain the discrimination correlations $p_2(\texttt{a,b}|\texttt{x},\diamond)$.
Based on the  independent and identically distributed assumption on experiment setups,  the state discrimination correlation is obtained as follows
\begin{eqnarray}
&&\!\!\!\!\!\!p_2(\texttt{a,b|x},\diamond)\\
=&&\!\!\!\!\!\!\tr\left[\bigotimes_{k=1}^N\left(M_{a_k|x_k}^{A_k}\otimes M_{b_k|y_k}^{B_{k0}B_k}(\tau^{A_kB_{k0}}\otimes\r^{B_1B_2...B_N})\right) \right],\nonumber
\end{eqnarray}
where $\texttt{y}=\diamond$ represents $\texttt{y}=(\diamond,\diamond,...,\diamond)$. 
If the correlations $p_1(\texttt{a,b,c|x,y,z})$ produce the maximal violation of the 3-CHSH inequalities $\b_{CHSH}$ in Eq. (\ref{ieq:bchsh}) and $\g_{CHSH_b}$ in  Eqs. (\ref{def:gchsh0})-(\ref{def:gchsh23}) between $N$ pairs of main party $B_k$ and auxiliary parties $A_k$ and $C_k$, then one can certify the following facts about the untrusted devices.  First the measurement performed on system $A_k$ are Pauli observables, and the auxiliary states $\tau^{A_kB_{k0}}$ are equivalent to the maximally entangled state $\proj{\Phi^0}$, up to some local unitaries. Second, when receiving the input $\texttt{y}=\diamond$, the main parties $B_k$ perform the Bell state measurement $\{\proj{\Phi^b}\}_{b=0}^3$, up to some local isometries. If the certification of experiment devices is fulfilled, i.e. the maximal violation occurs, then the state discrimination correlation is trustworthy.  It is derived by
\begin{eqnarray}
&&\!\!\!\!\!\!p_2(\texttt{a,b|x},\diamond)\nonumber\\
=&&\!\!\!\!\!\!\tr\bigg[\bigotimes_{k=1}^N\bigg(\pi_{a_k|x_k}^{A_k}\otimes \proj{\Phi^b}^{B_{k0}B_k}\\
&& \qquad\quad\times(\proj{\Phi^0}^{A_kB_{k0}}\otimes\r^{B_1B_2...B_N})\bigg) \bigg]\nonumber\\
=&&\!\!\!\!\!\!\frac{1}{4^N}\tr\left[\bigotimes_{k=1}^N (U_b^{B_k})^\dg\pi_{a_k|x_k}^{B_k}U_b^{B_k} \r^{B_1B_2...B_N}\right],
\end{eqnarray}
where the second equality follows from Eq. (\ref{eq:trphi}), and the unitaries $U_0=\bbI, U_1=\s_1, U_2=\s_2, U_3=\s_1\s_2$. 
One can see that the correlation $p_2(\texttt{a,b|x},\diamond)$ approaches to zero when the number of systems $N$ are large enough. So it can not be utilized for the state discrimination directly. From Eq. (\ref{eq:UbpiUb}), one has $U_{b_k}^\dg\pi_{a_k|x_k}U_{b_k}=\pi_{(-1)^{\delta_{(x,b)}+1}a|x}$. The referee turns to collect the statistics for each pair of systems $A_k$ and $B_k$ based on the category shown in Eqs. (\ref{eq:P2(a1)})-(\ref{eq:P2(a3)}). Each category  consists of four correlations that are equal to $p_2(a_k,0|x_k,\diamond)$ for system $k$, for $k=1,2,...,N$. The  correlations $ P_2(\texttt{a|x})$ can be obtained as follows
\begin{eqnarray}
\label{eq:P2N(ax)}
 P_2(\texttt{a|x})=&&\!\!\!\!\!\! 4^Np_2(\texttt{a},0|\texttt{x},\diamond)\\
 =&&\!\!\!\!\!\!
 \tr\left[\bigotimes_{k=1}^N \pi_{a_k|x_k}^{B_k} \r^{B_1B_2...B_N}\right].\nonumber
\end{eqnarray}
The correlations $P_2(\texttt{a|x})$ are used for the quantum state discrimination, and the measurement operators are the $N$-fold tensor product of Pauli observables. In fact, for any two different states $\ph_j$ and $\ph_k$ represented by Eq. (\ref{def:phj}), there exists at least one couple of coefficients such that
\begin{eqnarray}
	\label{eq:Sj1jN}
 S_{j_1j_2...j_N}^{(j)}\neq S_{j_1j_2...j_N}^{(k)}. 
\end{eqnarray}
Further if $\ph_j\neq \ph_k^*$, there exits at least one set of $j_1,j_2,...,j_N$ with $j_k=1,2$ satisfying Eq. (\ref{eq:Sj1jN}). For the convenience of state discrimination, one can choose the indices $m_k=1,2$ from the set such that $\abs{S_{m_1m_2...m_N}^{(j)}- S_{m_1m_2...m_N}^{(k)}}$ is maximum. If $\ph_j= \ph_k^*$, then the indices $m_k=1,2,3$ can be chosen such that $\abs{S_{m_1m_2...m_N}^{(j)}- S_{m_1m_2...m_N}^{(k)}}$ is maximum over all  $4^N$ coefficients in Eq. (\ref{def:phj}).
Then we choose the measurement strategy $\cM=\cM_N\in\{\s_{m_1}\otimes\s_{m_2}...\otimes \s_{m_N}\}$ in Eq. (\ref{def:normX_M}). It holds that for $t=j,k$,
\begin{eqnarray}
	\label{eq:sm1mn}
	S^{(t)}_{m_1m_2...m_N}=&&\!\!\!\!\!\!	\tr[(\s_{m_1}\otimes\s_{m_2}...\otimes\s_{m_N})\ph_t]
	\\
	=&&\!\!\!\!\!\!
	\tr\left[\bigotimes_{k=1}^N (\pi_{+1|m_k}-\pi_{-1|m_k}) \ph_t\right]\\
	=&&\!\!\!\!\!\!
	\sum_{a_1a_2...a_N}a_1a_2...a_N P^{\ph_t}(\texttt{a|m}).
\end{eqnarray}
 So the correlations $P_2(\texttt{a|m})$ are used to obtain the coefficient. Since $S_{m_1m_2...m_N}^{(j)}\neq S_{m_1m_2...m_N}^{(k)}$,  the states can be discriminated by this couple of coefficients.
On the other hand, the projectors of $\cM_N$ form an informationally complete set and hence they constitute a legitimate norm, by which the distance between the target states is derived by $d_{\cM_N}(\ph_j,\ph_k)=\frac{1}{2}\norm{\ph_j-\ph_k}_{\cM_N}$. 
One naturally obtains the following fact
\begin{lemma}
	\label{le:N qubit}
	The distance $d_{\cM_N}(\ph_j,\ph_k)>0$ for any different $N$-qubit states $\ph_j$ and $\ph_k$. 
\end{lemma}
It also demonstrates that the quantum state discrimination can be realized by the correlations $P_2(\texttt{a|x})$.

 As we have discussed above, if the states $\ph_j\neq\ph_k^*$ then the correlations $P(\texttt{a|x})$ with $x_k=1,2$ are suffice to realize the state discrimination device independently. Otherwise, if the states $\ph_j=\ph_k^*$ then some of the systems $B_k$ requires  a copy of trusted state $\eta=\proj{R}$ to determine the Pauli observable is $\s_3$ or $-\s_3$, and the states can be discriminated in a MDI manner.
 
 To conclude, for any different states $\ph_j$ and $\ph_k$, there exists appropriate inputs  $\texttt{x}$ and outputs $\texttt{a}$ such that  $P_2^{\ph_j}(\texttt{a|x})$ differ for different $j$ and $k$. We can distinguish them by observing the correspondence between  correlations and states given in Eq. (\ref{eq:sm1mn}). In quantum state tomography, $4^N-1$ coefficients in Eq. (\ref{def:phj}) remain to be determined. In this protocol, we can discriminate two states by a single coefficient  $S^{(t)}_{m_1m_2...m_N}$. Hence much less rounds of measurements is required in this protocol than tomography.

\section{guessing probability analysis for minimum-error discrimination}
\label{sec: mainguessing probability}
As we have shown in Sec. \ref{sec:discrimination}, the measurement operators in this protocol are restricted on  Pauli observables such that measurements do not depend on the target states and thus the DI requirement can be fulfilled. Next we quantify the influence of this restriction from the perspective of guessing probability for minimum error discrimination. We consider the discrimination of real  states measured by real Pauli observables, which can be realized in a DI scenario. 

Let $\{q_k,\ps_k\}_{k=1}^Q$ be an arbitrary ensemble with the prior knowledge, where the states are not generally orthogonal to each other. Since quantum states can not be discriminated perfectly, one naturally seeks a figure of merit, by which the measurement setting is optimized.  For instance, measurements are applied in a single-shot manner or performed repeatedly for independent and identically state distribution. 
Here we consider a well-known method for distinguishing quantum states, named minimum error discrimination \cite{bae2017Quantum}. For this method, the optimization over measurement devices requires  to seek for measurement $M$ with POVMs $\{M_k\}_{k=1}^Q$ such that the detection event on each $M_k$ leads to the state $\ps_k$ with minimum average error.  The guessing probability is then obtained by maximizing the success probability for correct guesses over all possible measurements,
\begin{eqnarray}
	\label{def:pguess}
p_G=\max_{M_k\geq 0, \sum_k M_k=\bbI}\sum_{k=1}^{Q}q_k\tr[M_k\ps_k].
\end{eqnarray}
The minimum average probability that one fails to guess correctly is $p_E=1-p_G$.

\begin{figure}[htp]
	\center{\includegraphics[width=8cm]  {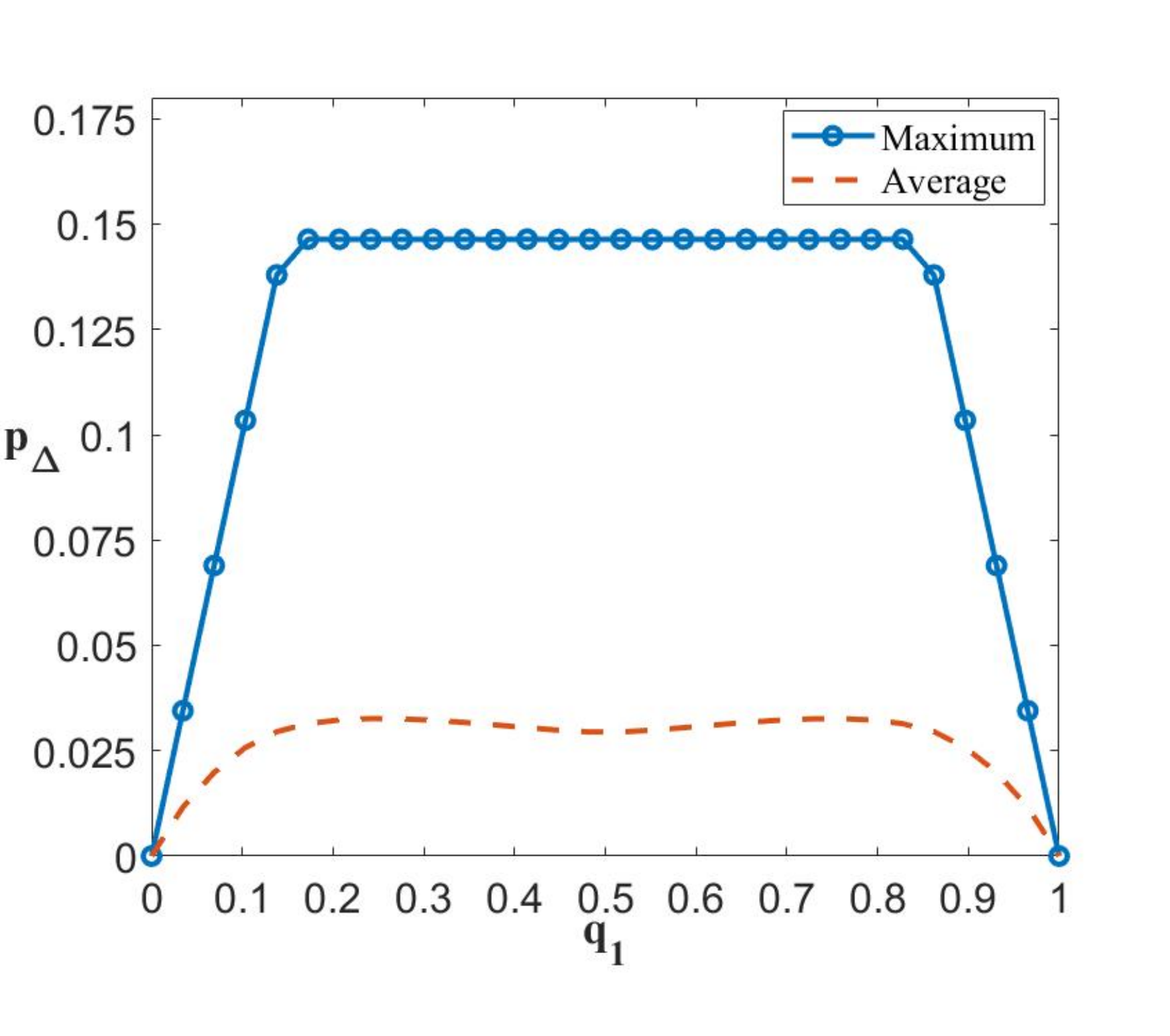}}
	\caption{Value of $\max_{\{\ps_1,\ps_2\}}p_{\Delta}$ and  $\mbox{avg}_{\{\ps_1,\ps_2\}}p_{\Delta}$  with respect to the probability $q_1$ for the ensemble $\{q_k,\ps_k\}_{k=1}^2$.}
	\label{fig:guessing1}
\end{figure}

\begin{figure}[htp]
	\center{\includegraphics[width=8cm]  {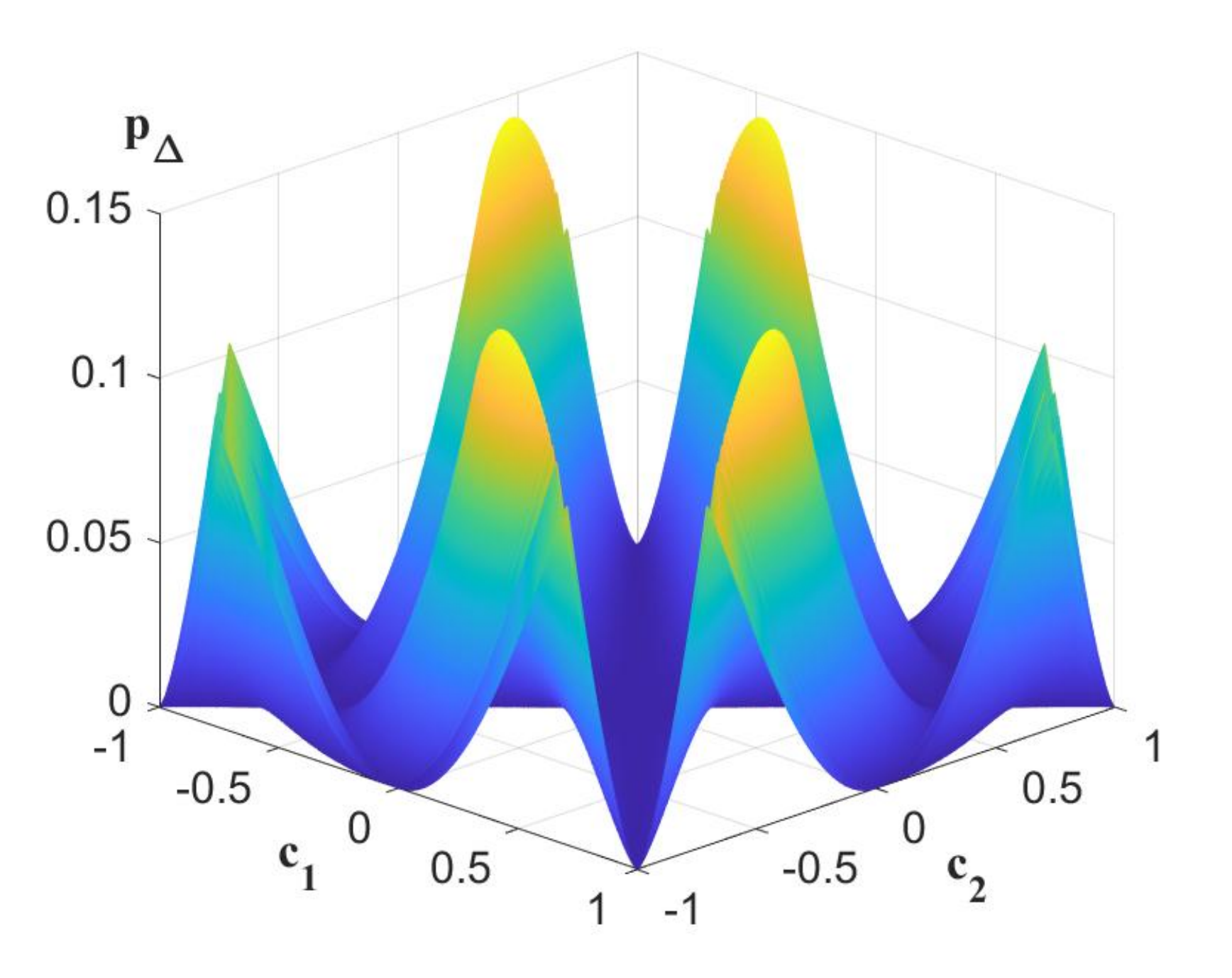}}
	\caption{ Value of $p_{\Delta}$ with respect to $c_k$ for the ensemble $\{q_k=1/2,\ps_k\}_{k=1}^2$ for $k=1,2$, where $\ps_k=\proj{\a_k}$ and $\ket{\a_k}=c_k\ket{0}+d_k\ket{1}$.}
	\label{fig:guessing2}
\end{figure}

As the only case where optimal discrimination is known without further assumptions \cite{Helstrom1969Quantum}, the discrimination between two quantum sates is considered below. By choosing $Q=2$ in Eq. (\ref{def:pguess}) and analysis shown in Appendix \ref{sec: guessing probability}, the guessing probability is then derived by 
\begin{eqnarray}
p_{G,1}=\frac{1}{2}+\frac{1}{2}\norm{q_1\ps_1-q_2\ps_2}_1,
\end{eqnarray}
where $\norm{\cdot}_1$ denotes the trace norm. This fact is known as the Helstrom bound \cite{Helstrom1969Quantum}. 
 By similar analysis as in Appendix \ref{sec: guessing probability}, the guessing probability for this protocol is measured by the norm $\norm{\cdot}_{M_1}$, which is derived by choosing $M=M_1\in\{\s_1,\s_2\}$ in Eq. (\ref{def:normX_M}).
 It is given by
 \begin{eqnarray}
 p_{G,2}=\frac{1}{2}+\frac{1}{2}\norm{q_1\ps_1-q_2\ps_2}_{M_1}.
 \end{eqnarray}
One can obtain that $p_{G,1}\geq p_{G,2}$ due to the restriction of measurements on the latter one. Further we show the value of 
\begin{eqnarray}
p_{\Delta}:= p_{G,1}-p_{G,2}	
\end{eqnarray}
 for an arbitrary ensemble $\{q_k,\ps_k\}_{k=1}^2$ by numerical analysis.
Traversing all possible combinations of real quantum states $\ps_1$ and $\ps_2$, we show  the average and maximum  value of $p_{\Delta}$ with respect to the probability $q_1$ (and thus $q_2=1-q_1$ is considered) in  FIG. \ref{fig:guessing1}.  The statistics show that $\max_{\{q_1,q_2\}}\mbox{avg}_{\{\ps_1,\ps_2\}}p_{\Delta}<0.033$, and $\max_{\{q_1,q_2\}}\max_{\{\ps_1,\ps_2\}}p_{\Delta}<0.146$. 
 In particular, when the states $\ps_1$ and $\ps_2$ occur with the same probability, i.e. $q_1=q_2=1/2$, the value of $p_{\Delta}$ with respect to $c_k$ is presented in FIG. \ref{fig:guessing2}, for $k=1,2$.
 The statistics imply that the  guessing probability is on average 3.3\% less than the optimal measurement discrimination  from the perspective of two arbitrary states. For the worst case, the guessing  probability in this protocol is maximally 14.6\% less than the optimal one. Note that $p_\Delta$ characterizes the difference of success rate for single-copy state discrimination. In this protocol,  the discrimination relies on the quantum correlations in Eq.  (\ref{eq:correP2(a|x)}). Multi-copy states are conventionally prepared, and the influence of $p_{\Delta}$ can be reduced practically.
 
From above analysis, one can obtain that  although the measurement operators are restricted on Pauli observables in this protocol,  the guessing probability is acceptable for quantum state discrimination.

\newpage
\section{Conclusion}
\label{sec:conclusion}
We have shown that  the quantum state discrimination can be realized  in a device independent scenario using tools  of self-testing  states and measurements. The  states can be discriminated credibly with the untrusted experiment devices by the correspondence between quantum correlations and states. Compared with the conventional  state discrimination, the security of this protocol is improved. In detail, we have shown that two arbitrary states can be discriminated in a DI manner when they are not conjugate with each other, while other states can be discriminated by the MDI way.   To fulfill the DI requirement, the measurements are restricted on Pauli observables, which form an informationally complete set. In order to  show the  influence of this restriction, we take all real states as an example and compare the guessing probability between Pauli measurement strategy and the optimal one. It is shown that the bias of guessing probability is less than 15\% on maximum and 0.33\% on average over all possible states. 

Many problems arising from this paper can be further explored. The DI discrimination of high-dimensional states can be realized with parallel self testing, and the structure of network may  follow the two-dimensional scenario. Noise robust analysis for this protocol can be established so that it can tolerate small amounts of experimental noise.

\section*{ACKNOWLEDGMENTS}
The authors were supported by the NNSF of China (Grant No. 11871089) and the Fundamental Research Funds for the Central Universities (Grant No. ZG216S2005).

\appendix
\section{Derivation of guessing probability for two-state discrimination}
\label{sec: guessing probability}
We review the derivation of guessing probability for two-state minimum-error discrimination.  It is measured by trace norm.  The analysis has been shown in Ref. \cite{bae2017Quantum}.

Choosing $Q=2$ in Eq. (\ref{def:pguess}), we have
\begin{eqnarray}
	p_{G,1}=&&\!\!\!\!\!\!\max_{M=\{M_1,M_2\}}(q_1\tr[M_1\ps_1]+q_2\tr[M_2\ps_2])\\
	:=&&\!\!\!\!\!\!\max_{M=\{M_1,M_2\}} \tr[K]\nonumber
\end{eqnarray}
with the operator $K=q_1M_1\ps_1+q_2M_2\ps_2$ depending on optimal POVMs $M_1$ and $M_2$. From $M_1+M_2=\bbI$, one has $K=q_2\ps_2+M_1X=q_1\ps_1-M_2X$ where $X=q_1\ps_1-q_2\ps_2$. By adding the above equations, it can be obtained that
\begin{eqnarray}
K=\frac{1}{2}(q_1\ps_1+q_2\ps_2)+\frac{1}{2}(M_1-M_2)X.
\end{eqnarray} 
 We set $M=M_1-M_2$, and the POVMs $M_{1/2}=(I\pm M)/2$ are both nonnegative. Then the guessing probability can be derived by the optimization of the measurement $M$ satisfying $-\bbI\leq M\leq \bbI$,
 \begin{eqnarray}
 p_{G,2}=&&\max_{M} \tr[K]=\frac{1}{2}+\frac{1}{2}\max_M\tr[MX]\\
 =&&\frac{1}{2}+\frac{1}{2}\norm{q_1\ps_1-q_2\ps_2}_1.\nonumber
 \end{eqnarray}

\bibliographystyle{unsrt}
\bibliography{DI}
\end{document}